\documentclass[prl, twocolumn, superscriptaddress, reprint]{revtex4-1}

\usepackage{amsmath}  % need for subequations
\usepackage{amsfonts}
\usepackage{graphicx}   % need for figures
\usepackage{hyperref}   % use for hypertext links, including those to external documents and UrLs

\usepackage{color}
\newcommand{\be}{\begin{enumerate}}
\newcommand{\ee}{\end{enumerate}}
\newcommand{\bi}{\begin{itemize}}
\newcommand{\ei}{\end{itemize}}

\newcommand{\xx}{\mathbf{x}}

\newcommand{\RNum}[1]{\uppercase\expandafter{\romannumeral #1\relax}}

\begin{document}

\title{Fracture in Sheets Draped on Curved Surfaces}
\author{Noah P. Mitchell}
\email{npmitchell@uchicago.edu}
\thanks{Corresponding author}
\affiliation{James Franck Institute and Department of Physics, The University of Chicago, Chicago, IL 60637, USA}
\author{Vinzenz Koning}
\affiliation{Instituut-Lorentz for Theoretical Physics, Universiteit Leiden, 2333 CA Leiden, The Netherlands}
\author{Vincenzo Vitelli}
\affiliation{Instituut-Lorentz for Theoretical Physics, Universiteit Leiden, 2333 CA Leiden, The Netherlands}
\author{William T. M. Irvine}
\email{wtmirvine@uchicago.edu}
\thanks{Corresponding author}
\affiliation{James Franck Institute and Department of Physics, The University of Chicago, Chicago, IL 60637, USA}
\affiliation{Enrico Fermi Institute, The University of Chicago, Chicago, IL 60637, USA}
\maketitle

\textbf{
Conforming materials to rigid substrates with Gaussian curvature --- positive for spheres and negative for saddles --- has proven a versatile tool to guide the self-assembly of defects such as scars, pleats~\cite{irvine_pleats_2010,bausch_grain_2003,bowick_two-dimensional_2009,vitelli_crystallography_2006,grason_universal_2013}, folds, blisters~\cite{holmes_draping_2010,hure_wrapping_2011}, and liquid crystal ripples~\cite{devries_divalent_2007}. 
Here, we show how curvature can likewise be used to control material failure and guide the paths of cracks.
In our experiments, and unlike in previous studies on cracked plates and shells~\cite{Slepyan_cracks_2002,Folias_Stresses_1965, amiri_phase-field_2014},
we constrained flat elastic sheets to adopt \textit{fixed} curvature profiles.
This constraint provides a geometric tool for controlling fracture behavior:
curvature can stimulate or suppress the growth of cracks, and steer or arrest their propagation. 
A simple analytical model captures crack behavior at the onset of propagation, while a two-dimensional phase-field model with an added curvature term successfully captures the crack's path. 
Because the curvature-induced stresses are independent of material parameters for isotropic, brittle media, our results apply across scales~\cite{rupich_soft_2014,Dusseault_Drilling_2004}.
}

Geometry on curved surfaces defies intuition: `parallel' lines diverge or converge as a consequence of curvature. 
As a result, when a thin material
conforms to such a surface, stretching and compression are inevitable~\cite{bowick_two-dimensional_2009}.
As stresses build up, the material can then respond by forming structures such as wrinkles or dislocations, 
which are themselves of geometric origin.
This interplay between curvature and structural response can result in universal behavior, independent of material parameters~\cite{hure_wrapping_2011,irvine_pleats_2010,bausch_grain_2003,vitelli_crystallography_2006,grason_universal_2013}.

A markedly different material response is to break via propagating cracks. 
While the use of curvature to control the morphology of wrinkles and defects in materials has been recently explored~\cite{irvine_pleats_2010,hure_wrapping_2011,bausch_grain_2003}, here we investigate the control of cracks by tuning the geometry of a rigid substrate. 
Can we design the underlying curvature of a substrate to steer paths of cracks in a material draped on that surface, thereby protecting certain regions?

\begin{figure}
 \includegraphics[width=\columnwidth]{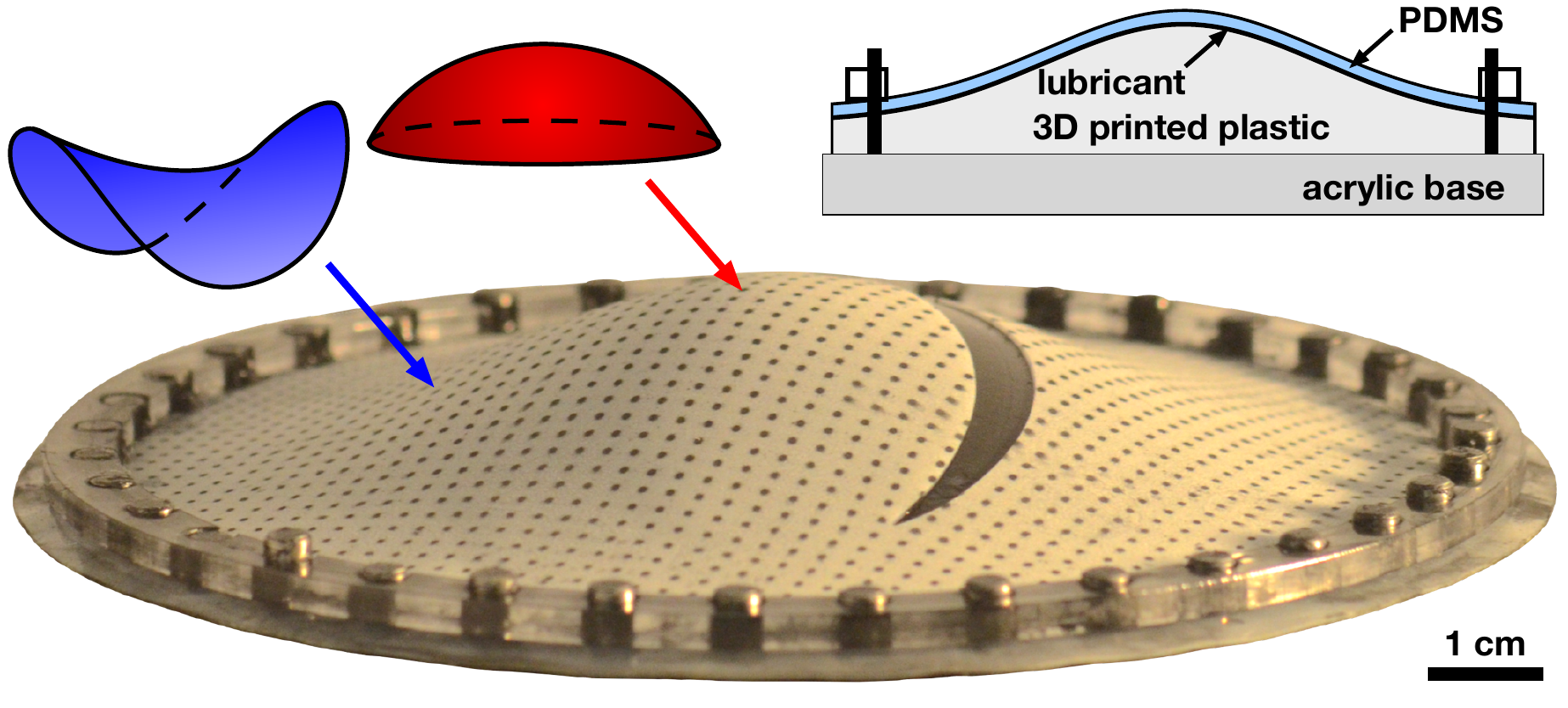}\\
\caption{Gaussian curvature --- positive for caps and negative for saddles --- governs the behavior of cracks. In the experimental setup, an initially flat PDMS sheet  conforms to a curved 3D printed surface. 
A small incision nucleates the crack.}
\label{fig1}
\end{figure}

To probe the effect of curvature on  cracks, we conform flat PDMS sheets (Smooth-On Rubber Glass II) to 3D-printed substrates (Fig.~\ref{fig1}). A lubricant ensures that the sheet conforms to the substrate while moving freely along the surface. 
We consider various geometries having positive and negative Gaussian curvature in both localized and distributed regions:  spherical caps, saddles, cones, and bumps. 
To begin, we focus on the bump as a model surface, as it is a common geometry containing regions of both positive and negative curvature.
A typical experimental run can be seen in \textit{Supplementary Videos 1-7}.
 We seed a crack by cutting a slit in the sheet, with a position and orientation of choice. 
By successive cuts, we increase their length until they exceed a critical length, known as the Griffith length~\cite{griffith_phenomena_1921, Rivlin_rupture_1953}, and propagate freely.

\begin{figure}[t]
\includegraphics[width=\columnwidth]{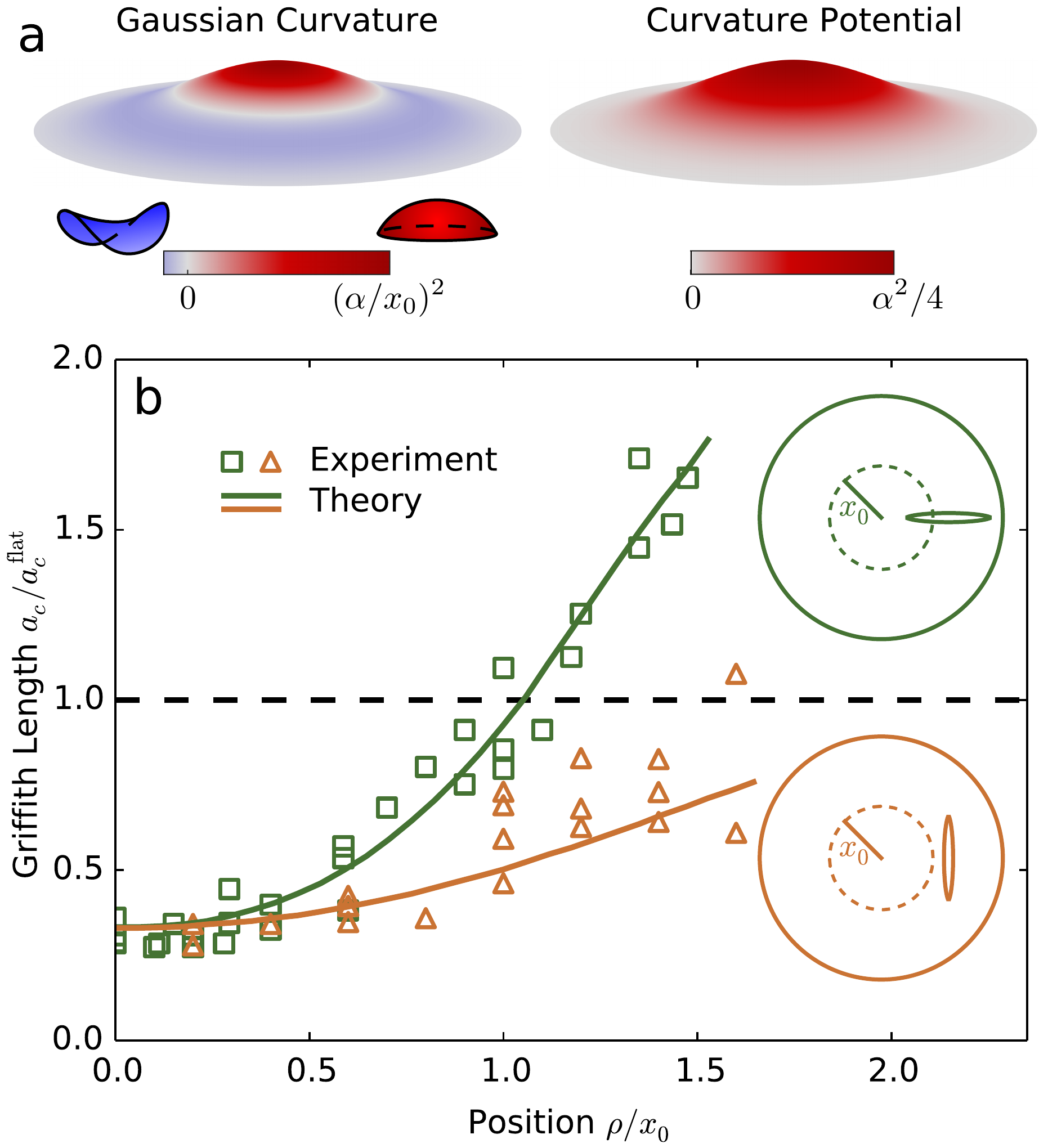}\\
\caption{
a) Gaussian curvature and curvature potential distributions for a bump with height profile $h(\rho) = \alpha x_0 \exp(-\rho^2/2x_0^2)$.
b) While the Griffith length for a crack in a flat sheet (dashed line) is nearly constant, curvature modulates the critical length of a seed crack. All samples shown had a 12 cm diameter ($2R$), an aspect ratio $\alpha=1/\sqrt{2}$, bump width $x_0=R/2.35$, and constant radial displacement $u_\rho/R=0.012$. 
}
\label{fig2}
\end{figure}

The Griffith length of a crack in a flat sheet is nearly independent of position and orientation. 
On our curved geometry, we find that this is not so. 
On the top of the bump, a shorter slit is necessary to produce a running crack, and on the outskirts of the bump (where the Gaussian curvature is negative), the behavior depends strongly on the orientation of the seed crack:
fracture initiation is suppressed for radial cracks, while the Griffith length for azimuthal cracks approaches that of the flat sheet (Fig.~\ref{fig2}b). Thus curvature can both stimulate and suppress fracture initiation, depending on the position and orientation of the seed crack.

\begin{figure}[h!]
\includegraphics[width=\columnwidth]{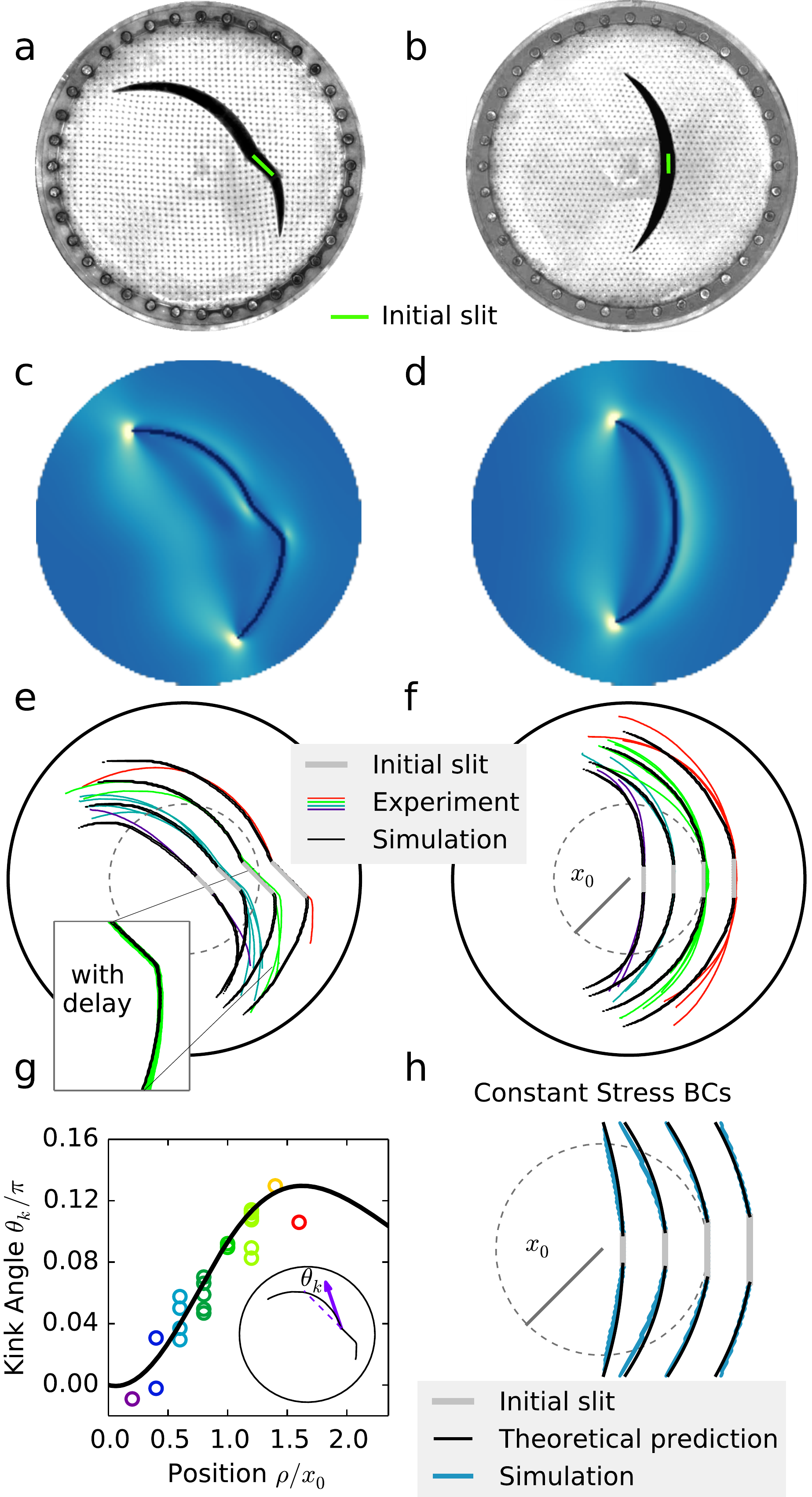}\\
\caption{
(a-b) Crack paths kink and curve around a bump. 
(c-d) Phase-field simulations of cracks on a bump, colored by the phase-modulated energy density so that broken regions are darkened.
(e-f) The phase-field crack path predictions (black solid curves) overlie the experimental paths (colored curves). 
(Inset) Introducing a time delay matching experiment for the right crack tip's propagation eliminates the discrepancy far from the bump.
(g) Analytical prediction (solid black curve) of the kink angle, $\theta_k$, overlies experimental results. 
(h) Analytical crack path predictions overlie simulations for free (constant stress) boundary conditions. 
All experiments and simulations had aspect ratio $\alpha=1/\sqrt{2}$ and bump width $x_0=R/2.35$, including the free boundary condition simulations.
}
\label{fig3}
\end{figure}
To relate these findings to the curvature distribution, we consider the stresses induced by curvature and their interaction with the crack tip.  
Stresses generated in the bulk of a material become concentrated near a crack tip.
In turn, a crack extends when the intensity of stress concentration exceeds a material-dependent, critical value~\cite{griffith_phenomena_1921,freund_dynamic_1990}. 
Expressed mathematically, in the coordinates of the crack tip ($r,\theta$), the stress in the vicinity of the tip takes the form
\begin{equation}\label{stress}
\sigma_{ij} = \frac{K_I}{\sqrt{2 \pi r}} f_{ij}^{I}(\theta) + \frac{K_{II}}{\sqrt{2\pi r}} f_{ij}^{II}(\theta),
\end{equation}
where $f_{ij}^{I,II}$ are universal angular functions~\cite{freund_dynamic_1990}. 
The factors $K_{I}$ and $K_{II}$ measure the intensity of tensile and shear stress concentration at the crack tip, respectively, and are known as stress intensity factors (SIFs). 
Thus, the Griffith length, $a_c$, is the length of the crack at which the intensity of stress concentration reaches the critical value, $K_c$.
In curved plates or sheets, the near-tip stress fields display the same singular behavior as in Eqn.~\ref{stress}~\cite{hui_williams_1998}, but the values of the SIFs are governed by curvature. 

Curving a flat sheet involves locally stretching and compressing the material by certain amounts at each point. 
According to the rules of differential geometry, this stretching factor, controlled by the field $\Phi$, is determined by an equation identical to the Poisson equation of electrostatics~\cite{vitelli_anomalous_2004}, with the Gaussian curvature, $G$, playing the role of a continuous charge distribution~\cite{bowick_two-dimensional_2009, vitelli_crystallography_2006}: 
\begin{equation}\label{harmonicPhi}
\nabla^2 \Phi(\xx) = - G(\xx).
\end{equation}
As the sheet equilibrates, its elasticity tends to oppose this mechanical constraint, giving rise to stress.  
The isotropic stress from curvature is then related to the potential via $\sigma^G_{kk}=E \Phi $, where $E$ is Young's modulus, and
the stress components are determined by integrals of the potential and boundary conditions (see Eqns.~25-26 of the \textit{Supplementary Information}).
Our study rests on a general geometric principle: 
positive (negative) curvature promotes local stretching (compression) of an elastic sheet, leading to the enhancement (suppression) of crack initiation.
Variations in the potential $\Phi$ steer the crack path, with the form of $\Phi$ determined nonlocally from the curvature distribution (see Eqn.~\ref{harmonicPhi} and Eqns. 39-41 of the \textit{Supplementary Information}).

For the bump, the curvature potential, $\Phi$, is large on the cap, where curvature is positive, and decays to zero as the negative curvature ring screens the cap (Fig.~\ref{fig2}a).
As $E\Phi$ is the isotropic stress, crack growth is stimulated where the potential is greatest --- on the cap of the bump, resulting in a small Griffith length there (Fig.~\ref{fig2}b).
Moving away from the cap, the potential decays, producing a stress asymmetry. 
This results in longer Griffith lengths with strong orientation dependence on the outskirts of the bump
(see Eqns. 39-41 of the \textit{Supplementary Information}). 
Fig.~\ref{fig2}b shows the theoretical results overlying the experimental data, with no fitting parameters.
We find that this minimal model is sufficient to capture the phenomenology of our system at the onset of fracture and provides correct qualitative predictions for longer cracks, even in the absence of symmetry. 

Curvature not only governs the critical length for fracture initiation, but also the direction of a crack's propagation.
For cracks inclined with respect to the bump, the cracks change direction as they begin to propagate, kinking at the onset of crack growth and curving around the bump, as shown in Fig.~\ref{fig3}a.
Cracks kink and curve towards the azimuthal direction because 
a decaying curvature potential, $\Phi(\rho)$, creates a local stress asymmetry: $\sigma^G_{\phi \phi} < \sigma^G_{\rho\rho}$. 
As a result, the crack relieves more elastic energy by deflecting towards the azimuthal direction. 
Analytical prediction of the kink angle, $\theta_k$, is made by selecting the direction of maximum hoop stress asymptotically near the crack tip (Eqn. 33 of the \textit{Supplementary Information}).
Fig.~\ref{fig3}g shows excellent agreement with experiment.

A purely analytical model is sufficient to capture the long-time behavior of the crack if the stress is fixed at the boundary (see~\ref{fig3}h).
This model extends the first order perturbation theory for slightly curved cracks developed by Cotterell and Rice~\cite{cotterell_slightly_1980} to curved sheets 
(see the section \textit{Perturbation Theory Prediction of Crack Paths} in the \textit{Supplementary Information}). 
As shown in Fig. 7 of the \textit{Supplementary Information}, the perturbation theory prediction is also increasingly accurate for constant displacement loading when the system size is large with respect to the crack.

\begin{figure}
\includegraphics[width=\columnwidth]{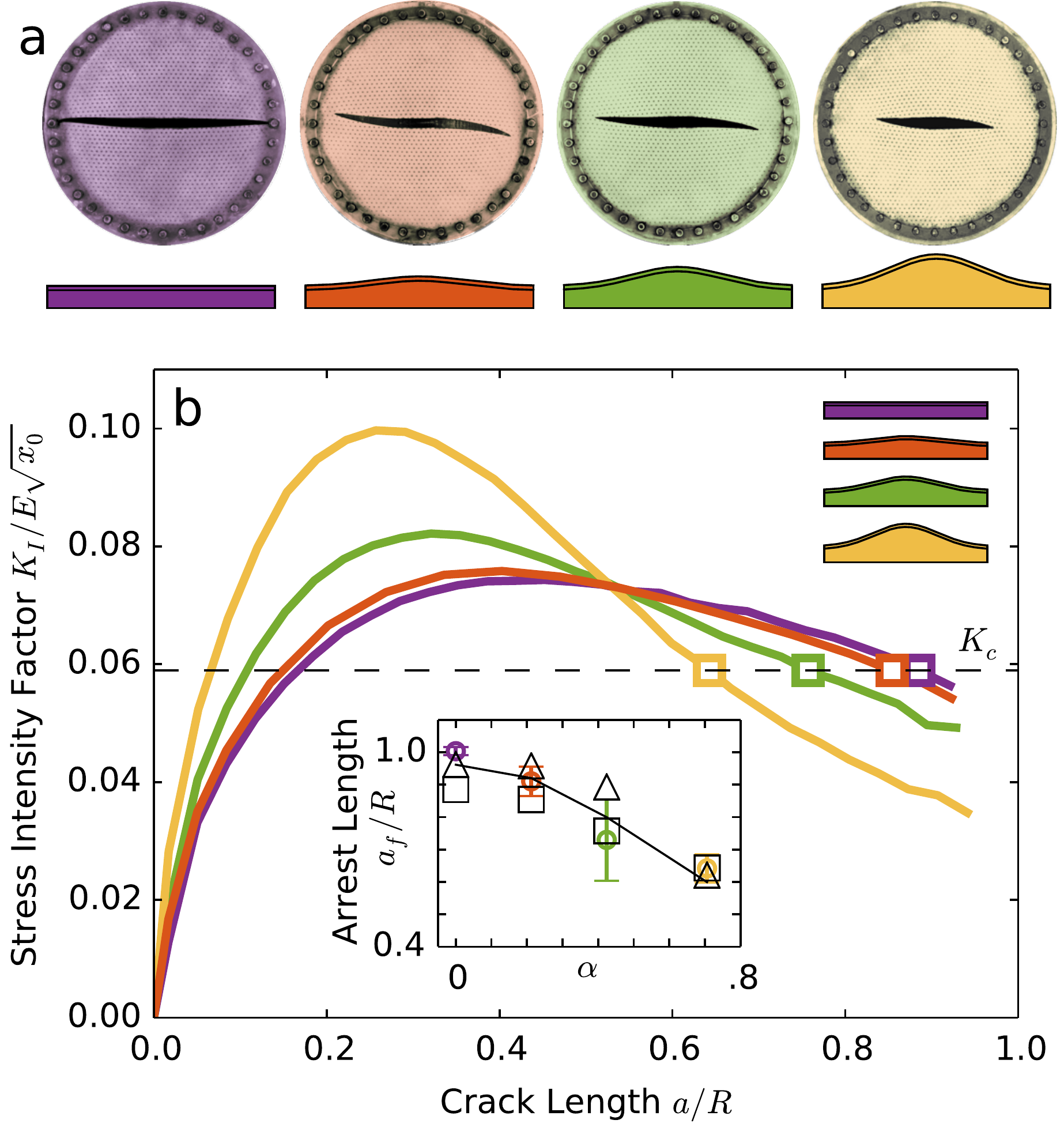}
\caption{
(a) Curvature arrests a center crack: as the aspect ratio of the bump increases while the initial stress at the boundary ($\sigma_{\rho\rho}(R) = 0.068\, E$) remains fixed, the final crack length decreases. 
(b) Simulations reveal that as the aspect ratio of the bump increases, the intensity of stress concentration falls below the critical value at progressively shorter crack lengths. 
\textit{Inset:} Final crack lengths from spring-lattice (squares) and phase-field simulations (triangles) mimic the arrest behavior seen in experiment (colored circles with error bars marking one standard deviation). 
The solid line is a guide to the eye.
}
\label{fig4}
\end{figure}

\begin{figure*}[t]
\includegraphics[width=\textwidth]{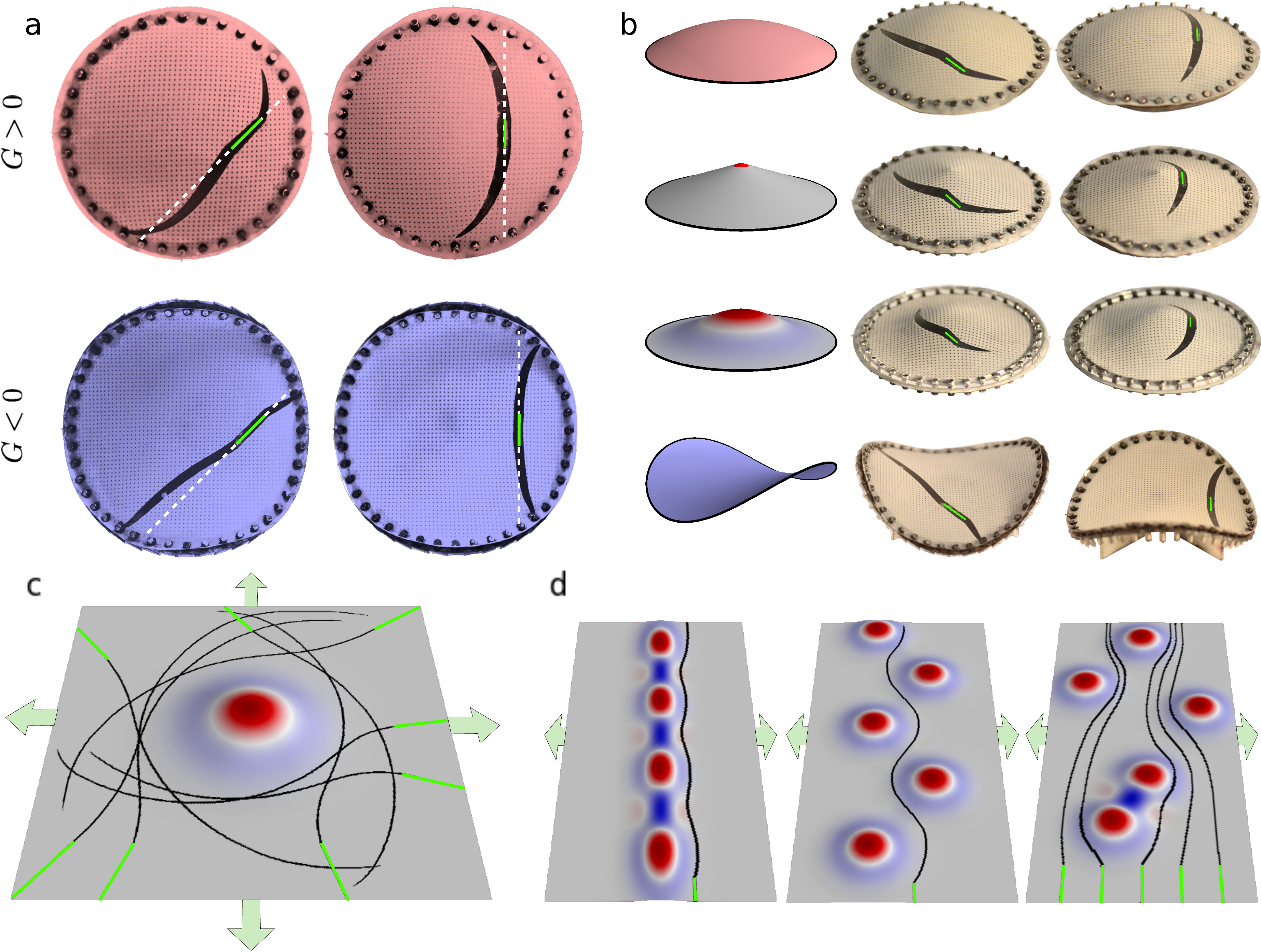}
\caption{
A crack's response to curvature exhibits universal behavior. 
(a) Inverting the sign of the curvature (red for positive, blue for negative) inverts the behavior of the crack, as shown by the contrasting crack paths on a $L=12$ cm spherical cap (top, $G=1/L^2$) and on a $L=$15 cm pseudospherical saddle (bottom, $G=-1/L^2$). 
Seed crack locations are marked in green.
(b) On spherical caps, cones, and bumps, the positive integrated curvature from the center to the crack's position directs cracks towards the azimuthal direction, while the negative curvature saddle inverts this behavior. 
(c) Further phase-field simulations demonstrate that curvature can protect a region of a material conformed to a bump (here under 3\% biaxial displacement) or (d) induce desired crack paths (here shown under 1.5\% uniaxial displacement). 
Final crack paths (black) for various initial slits (green) are overlaid to demonstrate that the bumps' central regions are protected. 
The results demonstrate that merely the addition of simple bumps offer a wide range of control, in experimentally realizable conformations.
}
\label{fig5}
\end{figure*}

For modest sample sizes with constant displacement boundary loading, however, a numerical approach is required.
To predict the curved fracture trajectories, we adapt the KKL phase-field model~\cite{karma_phase-field_2001,spatschek_phase_2011} to include curvature by incorporating the height profile of the substrate into the two-dimensional strain field~\cite{Nelson_fluctuations_1987}. 
This numerical model treats local material damage as a scalar field that evolves if there is both sufficient elastic energy density and a local gradient in the field (see \textit{Supplementary Information}). 
As depicted in Fig.~\ref{fig3}c and d, these conditions are met at the tip of a propagating crack.
This model captures the full crack paths, as shown by the black curves overlying experimental results in Fig.~\ref{fig3}e and~\ref{fig3}f.

A systematic deviation in the extensions of the crack tips further from the bump is evident in Fig.~\ref{fig3}e.
In the experiments, the tip closer to the bump begins its advance first, and the dynamics of the tip are not purely quasistatic. 
In the phase-field simulation, simply suppressing the tip further from the bump for a short time until the near tip has reached a distance matching experiment eliminates this deviation, as shown in the inset of Fig.\ref{fig3}e (see the \textit{Phase-Field Model} section of the \textit{Supplementary Information} for details).

Having seen how curvature affects the initiation and propagation of cracks, we now turn our attention to the ability of curvature to arrest cracks.
As seen previously in Fig.~\ref{fig3}, curved cracks can terminate before reaching the sample boundary.
We find, moreover, that curvature can arrest cracks even for cases in which the path is undeflected, as shown in Fig.~\ref{fig4}.
In flat sheets, center cracks propagate all the way to the boundary, but if we introduce a bump while holding the initial stress at the boundary fixed, the final crack length decreases.

From the decaying isotropic stress profile, we can infer that curvature generates azimuthal compression, halting the crack's advance.
Using our phase-field model, we indeed find that increasing the aspect ratio of the bump lowers the intensity of stress concentration for larger crack lengths (Fig.~\ref{fig4}b). 
A fully 3D spring network simulation using finite element methods provides additional confirmation (open squares in Fig.~\ref{fig4}b). 
Thus, curvature decreases the final crack length, despite promoting crack initiation on top of the bump.

Curvature's influence on the propagation of cracks that we have investigated on the bump is not peculiar to that surface.
As shown in Fig.~\ref{fig5}, we demonstrate this generality by testing a number of additional surfaces, including spherical caps (uniform $G>0$), cones ($G=G_0 \,\delta(\xx)$), and pseudospherical saddles (uniform $G<0$). 
A region of positive curvature, such as the tip of a cone, 
locally stimulates crack growth near the region, but also guides cracks around that region. 
Conversely, negative curvature of a saddle suppresses crack growth and orients cracks away from the center (see Fig.~\ref{fig5} and \textit{Supplementary Information}).
Thus an opposite curvature source induces an opposite response: the behavior of cracks is tunable by engineering the curvature landscape.

In Fig.~\ref{fig5}c and d, we demonstrate the robustness of curvature's effects by considering samples without azimuthal symmetry using the phase-field model. 
Here, we use a bump to protect a central region from incoming cracks of various orientations, to produce oscillating cracks, and to focus and diverge possible crack paths.
For the geometries of Fig.~\ref{fig5}d, a somewhat reduced critical stress intensity factor compared to our experimental material prevents crack arrest. 
Though the stress is highest on top of a bump, these regions are protected from approaching cracks (see \textit{Supplementary Video 8}).

%We have shown that conforming flat sheets to rigid, curved substrates can direct fracture initiation, guide crack trajectories, and arrest cracks. 
The use of substrate curvature to control fracture morphology differs from using existing cracks or inclusions in that our method requires no introduction of pre-existing structure into the fracturing sheets~\cite{ghelichi_modeling_2015,cheeseman_interaction_2000}.
%The resulting fracture behavior can be understood in terms of a simple model, which relates curvature to the generation of nonlocal stresses.
For brittle sheets with isotropic elasticity, curvature-induced stresses are independent of material parameters and only dependent on geometry. 
Therefore, our results represent the effects of substrate curvature on fracture morphology for a wide range of materials, 
with potential implications for thin films, monolayers~\cite{rupich_soft_2014,yuk_high-resolution_2012}, 
geological strata such as near salt diapirs~\cite{price_analysis_1990,Dusseault_Drilling_2004}, and stretchable electronics~\cite{rogers_materials_2010}.
Since the results are based on the modulations of the material's metric, they should also apply beyond conformed sheets, with metrics engineered by other methods
--- for instance, temperature gradients \cite{yuse_transition_1993} or differential swelling~\cite{sharon_mechanics_2010}. 

\section{Acknowledgements}
The authors thank Efi Efrati, Hridesh Kedia, Dustin Kleckner, Michelle Driscoll, Sid Nagel, Tom Witten, and Ridg Scott for interesting discussions and Jacob Mazor for assistance with some supplementary experiments. Some simulations were carried out on the Midway Cluster provided by the University of Chicago Research Computing Center.
We acknowledge the Materials Research and Engineering
Centers (MRSEC) Shared Facilities at The University of Chicago
for the use of their instruments. 
This work was supported by the National Science Foundation MRSEC Program at The University of Chicago (Grant DMR-1420709) and a Packard Fellowship.
V.K. and V.V. acknowledge funding from FOM and NWO.

\section{Author contributions}
W.T.M.I. and V.V. initiated this study. N.P.M. and W.T.M.I. designed experiments. N.P.M. performed and analyzed the experiments and simulations. N.P.M. and V.K. constructed the analytical model. All authors interpreted the data. N.P.M., V.V., and W.T.M.I wrote the manuscript.

\section{Code availability}
Custom python codes for phase-field model simulations and analytical crack trajectories are available at https://github.com/irvinelab/fracture, including detailed documentation.

\end{document}